\documentclass[12pt, a4paper]{article}
\usepackage[left=1.5cm, right=1.5cm, top=2cm, bottom=2cm]{geometry}
\usepackage[square,numbers]{natbib}

\usepackage{amsmath,amssymb,amsthm}
\usepackage{graphicx}
\usepackage{hyperref}
\usepackage{algorithm,algpseudocode}
\usepackage{booktabs}
\usepackage{siunitx}

\title{Explicit asymptotics of coupling matrix elements for central potentials in the hyperspherical harmonics expansion method}
\author{Emile  Meoto$^1$  \and Mantile L. Lekala$^2$ }
\date{$^1$ Department of Physics, University of Buea, P.O. Box 63 Buea, South West Region, Cameroon, \\
Email: Meoto.Emile@ubuea.cm or EmileMeoto@aims.ac.za \\
$^2$ Department of Physics, University of South Africa, Florida Park, P.O. Box 392, South Africa, \\
Email: lekalml@unisa.ac.za \\
02 March 2026}
\begin{document}

\maketitle

\begin{abstract}
The analytic structure and asymptotic behavior of channel-coupling potentials in three-body systems are investigated within the framework of the hyperspherical harmonics expansion method. The coupling between different Jacobi partitions is expressed using Raynal--Revai transformation coefficients and a reduced hyperangular integral that contains the two-body interaction. For central potentials, this integral is factorised into geometric and dynamical components. Explicit asymptotic scaling laws are derived for the hyperradial coupling strength in the limit of large hyperradius $\rho \to \infty$ for representative nuclear potentials: Gaussian, Yukawa, and Woods--Saxon potentials (short-range), and Coulomb potential (long-range). These short-range potentials are found to exhibit an algebraic decay $\propto \rho^{-(2\ell_{\eta_i}+3)}$, where $\ell_{\eta_i}$ is the orbital angular momentum of the interacting pair. This decay is shown to lead to efficient asymptotic decoupling of hyperspherical channels. In contrast, the Coulomb interaction yields couplings that decay only as $1/\rho$, indicating persistent channel coupling at large distances and explaining the slow convergence of hyperspherical expansions for charged systems. These results provide a quantitative basis for truncating the hyperradial domain, for example, in choosing a matching radius for scattering calculations or an upper integration limit in bound-state problems.
\end{abstract}

\section{Introduction}

The three-body problem lies at the heart of many fundamental questions in nuclear, atomic, and molecular physics. Despite its long history, obtaining accurate solutions remains challenging due to the non-separable nature of the interactions and the complexity of the coupled-channel equations that arise in many-body descriptions. Among the various methods developed to treat three-body systems, the hyperspherical harmonics (HH) expansion has emerged as a powerful and systematically improvable approach. In this method, the wave function is expanded in a complete set of hyperangular functions, reducing the problem to a system of coupled hyperradial equations. The efficiency and convergence of the HH expansion, however, depend critically on the behavior of the coupling potentials between different hyperspherical channels, particularly in the asymptotic region of large hyperradius.

A practical implementation of the HH expansion method requires a truncation of the expansion. The choice of truncation is guided by the decay properties of the channel-coupling potentials: if these couplings vanish sufficiently rapidly with increasing hyperradius, the channels decouple asymptotically, and a limited number of terms can capture the essential physics. Conversely, slowly decaying couplings necessitate the inclusion of many channels, complicating numerical computations and limiting the applicability of the method. Understanding the analytic structure and asymptotic scaling of these couplings is therefore essential for both interpreting computational results and designing efficient numerical schemes.

In this paper, a detailed investigation of the asymptotic behavior of channel-coupling potentials within the HH formalism for three-body systems is presented. We focus on representative nuclear interactions: the Gaussian, Yukawa, and Woods-Saxon potentials as prototypes of short-range interactions, and the Coulomb potential as a classic long-range interaction. Our analysis not only elucidates the origin of coupling in three-body systems but also offers quantitative criteria for truncating hyperspherical expansions in practical calculations. The explicit asymptotic formulas derived here can be used to assess the importance of distant couplings and to develop more efficient numerical strategies for solving the three-body Faddeev equations, both for bound states and scattering processes. 

\section{Hyperspherical coordinates}

Two of the most commonly used hyperspherical coordinates in nuclear physics are Delves coordinates and Smith-Whitten coordinates. Delves coordinates are applied in this paper. For a three-body problem, Delves hyperspherical coordinates consist of one hyperradius, one hyperangle and four polar angles. After separation of the centre-of-mass motion, the internal dynamics of a three-body system is described by two mass-scaled Jacobi vectors that are given in spectator notation by $(\vec{\eta}_i, \vec{\lambda}_i)$. From this six-dimensional Jacobi coordinates, six-dimensional Delves hyperspherical coordinates ($\rho$, $\theta_i$, $\nu_{\eta_i}$,  $\omega_{\eta_i}$, $\nu_{\lambda_i}$, $\omega_{\lambda_i}$) are constructed. The coordinate \(\rho \in [0, \infty)\) is the hyperradius, a collective coordinate representing the overall size of the three-body system. It is defined as \(\rho = \sqrt{\eta_i^2 + \lambda_i^2}\) and is invariant under changes of Jacobi coordinate sets. This invariance follows from the orthogonality of the Jacobi transformation between partitions as illustrated in \cite{meo2026}. The coordinate $\theta_i \in [0, \pi/2]$ is the hyperangle and is defined through the parametrisation 

\begin{align}
\eta_i= \rho \sin \theta_i \\
\lambda_i = \rho \cos \theta_i
\end{align}

The pair of coordinates $\Omega_{\eta_i}=(\nu_{\eta_i},  \omega_{\eta_i})$ are spherical polar angles associated with the Jacobi vector $\vec{\eta}_i$ while $\Omega_{\lambda_i}=(\nu_{\lambda_i}, \omega_{\lambda_i})$ are spherical polar angles associated with $\vec{\lambda}_i$. This means $(\eta_i, \nu_{\eta_i},  \omega_{\eta_i})$ and $(\lambda_i, \nu_{\lambda_i}, \omega_{\lambda_i})$ form separate spherical coordinate systems. While the hyperradius defines the overall size of the system, these angular coordinates describe the internal structure.

Two-body wavefunctions in Jacobi coordinates $\psi_i(\vec{\eta}_i, \vec{\lambda}_i)$ are written as $\psi_i(\rho, \Omega^i_5)$ in hyperspherical coordinates. This wavefunction is separated as follows:

\begin{align}
\label{eq:expansion}
    \psi_{i}^{J\alpha_{i}}(\rho,\Omega_{5}^{i})=\sum_{K_{i}}\frac{\chi_{K_{i}}^{i,J\alpha_{i}}(\rho)}{\rho^{5/2}}\mathcal{Y}_{K_{i}}^{\ell_{\eta_{i}}\ell_{\lambda_{i}}}(\Omega_{5}^{i})
\end{align}

where \(\chi^{i,J\alpha_i}_{K_i}(\rho)\) is the hyperradial wavefunction for channel \(i\), with total angular momentum  of the three-body system \(J\), channel quantum numbers \(\alpha_i\) (e.g., spin, orbital angular momentum coupling), and hyperangular momentum \(K_i\). The superscript \(i\) indicates the spectator particle (Jacobi set \(i\)). $\mathcal{Y}_{K_{i}}^{\ell_{\eta_{i}}\ell_{\lambda_{i}}}(\Omega_{5}^{i})$ is the hyperspherical harmonic for channel $i$, with orbital angular momenta $\ell_{\eta_{i}}$ and $\ell_{\lambda_{i}}$. $\Omega_{5}^{i}$ represents the five angles ($\theta_i$, $\nu_{\eta_i}$,  $\omega_{\eta_i}$, $\nu_{\lambda_i}$, $\omega_{\lambda_i}$) for the $i$-th Jacobi set.

In the hyperspherical harmonics expansion method for solving the three-body problem, the wavefunction expansion in Eq. \eqref{eq:expansion} is substituted into the Faddeev equations expressed in Jacobi coordinates. This equation is then projected on a complete hyperspherical harmonics basis. The details of this projection are outlined in \citep{meo2026}. This projection yields a system of coupled hyperradial equations:

\begin{equation}
\left[-\frac{\hbar^{2}}{2m}\frac{d^{2}}{d\rho^{2}}+L_{K_{i}}-E\right] \chi^{i,J\alpha_{i}}_{K_{i}}(\rho)+\sum_{n,K_{n}}V^{in}_{K_{i}K_{n}}(\rho)\chi^{n,J\alpha_{n}}_{K_{n}}(\rho)=0 ,
\end{equation}

where \(\hbar\) is the reduced Planck’s constant, \(m\) is a reference mass introduced to scale the Jacobi coordinates (in nuclear physics, this is often taken as the neutron mass \(939.6\,\text{MeV}/c^2\)), \(E\) is the total energy of the three-body system (internal energy, since the centre-of-mass motion has been separated out) and \(n\) is an index labelling different Faddeev components (channels) corresponding to different Jacobi coordinate sets (\(n = i, j, k\)). \(L_{K_i}\) is a centrifugal barrier, given by  

    \begin{align}
            L_{K_i} = \frac{\hbar^2 \left(K_i(K_i+4) + \frac{15}{4}\right)}{2m\rho^2}.
    \end{align}
     
As elucidated in \citep{meo2026}, this centrifugal barrier arises from the hyperradial part of the kinetic energy operator and includes contributions from the grand angular momentum operator $\mathbf{\Lambda}^2(\Omega^i_5)$. \(V^{in}_{K_i K_n}(\rho)\) is the coupling potential between hyperradial channels \((i, K_i)\) and \((n, K_n)\), defined as  

    \begin{align}
    \label{eq:define_coupling}
            V^{in}_{K_i K_n}(\rho) = \int \left[\mathcal{Y}^{\ell_{\eta_i}\ell_{\lambda_i}}_{K_i}(\Omega_5^i)\right]^* V_i(\rho, \Omega_5^i) \, \mathcal{Y}^{\ell_{\eta_n}\ell_{\lambda_n}}_{K_n}(\Omega_5^n) \, d\Omega_5^i.
    \end{align}

where \(V_i(\rho, \Omega_5^i)\) is the interaction potential in the \(i\)-th partition and $d\Omega_5^i = \sin^2\theta_i \cos^2\theta_i \, d\theta_i \, d\Omega_{\eta_i} \, d\Omega_{\lambda_i}$. The full 6D volume element transformation from mass-scaled Jacobi coordinates to Delves hyperspherical coordinates is $d^3\vec{\eta}_i \, d^3\vec{\lambda}_i = \rho^5 \, d\rho \, d\Omega_5^i$. For realistic nuclear interactions, the integral in Eq. \eqref{eq:define_coupling} must generally be evaluated numerically (see, e.g., \citep{tho2004}). While Eq. \eqref{eq:define_coupling} provides a formal definition of the coupling potential, its direct evaluation is complicated by the presence of hyperspherical harmonics belonging to different Jacobi partitions. A practical evaluation therefore requires expressing all hyperspherical harmonics in a common coordinate system. This transformation is accomplished by the Raynal-Revai coefficients.

\section{Transformation through Raynal--Revai coefficients}

The Raynal-Revai coefficients  \citep{ray1970, ray1973} provide a unitary transformation between hyperspherical harmonics corresponding to different Jacobi coordinate sets. They allow us to express harmonics in the $n$-partition as an expansion over harmonics in the $i$-partition:

\begin{equation}
\mathcal{Y}^{\ell_{\eta_n}\ell_{\lambda_n}}_{K_n}(\Omega_5^n) = \sum_{\ell_{\eta_i},\, \ell_{\lambda_i}} RR\langle \ell_{\eta_i} \ell_{\lambda_i} | \ell_{\eta_n} \ell_{\lambda_n} \rangle_{\mathrm{K_nL}} \; \mathcal{Y}^{\ell_{\eta_i}\ell_{\lambda_i}}_{K_n}(\Omega_5^i),
\label{eq:RR-expansion}
\end{equation}

where \(RR\langle \ell_{\eta_i} \ell_{\lambda_i} | \ell_{\eta_n} \ell_{\lambda_n} \rangle_{\mathrm{K_nL}}\) are the Raynal--Revai coefficients. Substituting this expansion into the equation for the coupling potential yields

\begin{equation}
\label{eq:coupling-with-RR}
V^{in}_{K_i K_n}(\rho)
=
\sum_{\ell_{\eta_i}',\,\ell_{\lambda_i}'}
RR\left\langle
\ell_{\eta_i}' \ell_{\lambda_i}'
\,\middle|\,
\ell_{\eta_n} \ell_{\lambda_n}
\right\rangle_{\mathrm{K_nL}}
\int
\left[
\mathcal{Y}^{\ell_{\eta_i}\ell_{\lambda_i}}_{K_i}(\Omega_5^{i})
\right]^{*}
\, V_i(\rho,\Omega_5^{i}) \,
\mathcal{Y}^{\ell_{\eta_i}'\ell_{\lambda_i}'}_{K_n}(\Omega_5^{i})
\, d\Omega_5^{i}.
\end{equation}

The integral is now entirely within the same Jacobi set \(i\). This integral is a matrix element of the potential \(V_i\) between hyperspherical harmonics of the same coordinate system, which can be evaluated using angular momentum algebra and numerical integration. 

The goal of this project is to explore the asymptotic behaviour of this coupling for a few classes of potentials that are frequently encountered in nuclear physics. We shall focus on three short-range potentials (Gaussian, Yukawa and Woods-Saxon potentials) and one long-range potential (Coulomb potential).

\section{Central potentials}

The nuclear force is not purely central: it contains tensor components, spin–orbit coupling, and depends on the relative spins and isospins of the nucleons. While purely central potentials can reproduce binding energies, many other observables can only be described through the inclusion of non-central forces, even in simple systems such as the deuteron and the triton. For example, a central potential alone cannot account for the quadrupole moment and level splittings in the deuteron or for detailed nucleon-nucleon scattering data.

Despite these limitations, central potentials are frequently used in few-body calculations because they can reproduce binding energies and provide valuable benchmarks. Typical applications include three-nucleon systems such as the triton and three-boson systems such as the $\alpha+\alpha+\alpha$ problem. They are also widely employed when different few-body techniques are compared, aiding methodological development and validation. Central potentials therefore serve as a foundational starting point in few-body nuclear physics, before being supplemented or replaced by more realistic interactions that include non-central terms. Crucially, the central component often dominates the spatial structure of the wave function, while non-central forces act as refinements rather than fundamentally altering the interaction. At very low energies, the interaction is dominated by the $s$-wave, which is primarily central.

Commonly used central potentials include the Malfliet–Tjon potential in three-nucleon calculations \citep{mal1969, fil2023} and the Ali–Bodmer potential in three-$\alpha$ systems \citep{sof2013, fil2022}. In some three-body models, combinations of central potentials are employed. For example, in \citep{fil2007} the Ali–Bodmer $\alpha\alpha$ interaction and a central $\alpha\Lambda$ potential were used to describe ${}^{9}_{\Lambda}\mathrm{Be}$ as a $\Lambda+\alpha+\alpha$ three-body system.

An additional motivation for studying central potentials is that they are often used to simulate realistic interactions, owing to their rapid convergence in few-body calculations. Realistic nuclear and hypernuclear potentials typically involve many operator components, making them computationally demanding. Central-only representations are therefore constructed to capture the essential short-range behavior while remaining numerically efficient. For instance, in \citep{meo2019, meo2020} the Nijmegen Soft-Core NSC97f $\Lambda N$ interaction was simulated using inverse scattering theory to obtain the GLM-YN0 central potential. Similarly, lattice QCD (HAL QCD) potentials are routinely fitted by purely central forms for use in three-body calculations \citep{etm2022, iri2019}.

In non-relativistic atomic physics, the Coulomb interaction acts between all charged particle pairs and is a central force. Well-known three-body atomic systems include the helium atom, muonic helium, negative ions such as H$^-$ and Ps$^-$, and electron–atom scattering systems.

In atomic and molecular few-body systems composed of neutral particles, the dominant low-energy two-body interaction is the van der Waals force. In its leading approximation, this interaction is isotropic and depends only on the interparticle separation, and is therefore central. A notable three-body example is the helium trimer, consisting of three ${}^4$He atoms, which supports an Efimov excited state bound by van der Waals interactions.

\subsection{Structure of potential coupling for central potentials}

In the \(i\)-th Jacobi partition, the two-body potential depends only on the relative coordinate between particles \(j\) and \(k\). In Jacobi coordinates \((\vec{\eta}_i, \vec{\lambda}_i)\), this relative coordinate is simply \(\vec{\eta}_i\). Therefore, a central potential is of the form

\begin{equation}
V_i(\vec{\eta}_i) = V_i(|\vec{\eta}_i|).
\end{equation}

In hyperspherical coordinates, using \(\eta_i = \rho \sin\theta_i\), this becomes:

\begin{equation}
V_i(\rho, \theta_i) = V_i(\rho \sin\theta_i).
\end{equation}

A central potential is independent of the four spherical angles \((\nu_{\eta_i}, \omega_{\eta_i}, \nu_{\lambda_i}, \omega_{\lambda_i})\). This property of central potentials greatly simplifies the angular integration for the coupling potential. Eq. \ref{eq:coupling-with-RR} may be written as

\begin{equation}
\label{eq:V_in_factorized}
V^{in}_{K_i K_n}(\rho)
=
\sum_{\ell_{\eta_i}',\,\ell_{\lambda_i}'}
RR\left\langle
\ell_{\eta_i}' \ell_{\lambda_i}'
\,\middle|\,
\ell_{\eta_n} \ell_{\lambda_n}
\right\rangle_{\mathrm{K_nL}}
\,
V^{\ell_{\eta_i},\ell_{\lambda_i};\,\ell_{\eta_i}',\ell_{\lambda_i}'}_{K_iK_n}(\rho),
\end{equation}

where the channel coupling integral \(V\) is:

\begin{equation}
\label{eq:I_general}
V^{\ell_{\eta_i},\ell_{\lambda_i};\,\ell_{\eta_i}',\ell_{\lambda_i}'}_{K_iK_n}(\rho)
=
\int
\left[
\mathcal{Y}^{\ell_{\eta_i}\ell_{\lambda_i}}_{K_i}(\Omega_5^{i})
\right]^{*}
\, V_i(\rho \sin\theta_i) \,
\mathcal{Y}^{\ell_{\eta_i}'\ell_{\lambda_i}'}_{K_n}(\Omega_5^{i})
\, d\Omega_5^{i}.
\end{equation}

The five-dimensional angular volume element $d\Omega^i_5$ factorises as:
\begin{equation}
d\Omega^i_5 = \sin^2\theta_i \cos^2\theta_i \; d\theta_i \; d\Omega_{\eta_i} \; d\Omega_{\lambda_i},
\end{equation}
where  
$d\Omega_{\eta_i} = \sin\nu_{\eta_i} \, d\nu_{\eta_i} \, d\omega_{\eta_i}$ is the solid angle for $\vec{\eta}_i$, and  
$d\Omega_{\lambda_i} = \sin\nu_{\lambda_i} \, d\nu_{\lambda_i} \, d\omega_{\lambda_i}$ is the solid angle for $\vec{\lambda}_i$.

\subsection{Hyperspherical harmonics}

Hyperspherical harmonics are constructed by coupling spherical harmonics \(Y^{\ell_{\eta_i} m_{\eta_i}}\) and \(Y^{\ell_{\lambda_i} m_{\lambda_i}}\) through the hyperangle \(\theta_i\): 

\begin{align}
\label{eq:hyperangular}
\mathcal{Y}^{\ell_{\eta_i} \ell_{\lambda_i}}_{K_i}(\Omega^i_5)= \phi_{K_i}^{\ell_{\eta_i}, \ell_{\lambda_i}}(\theta_i) \left[ Y^{\ell_{\eta_i} m_{\eta_i}} \otimes Y^{\ell_{\lambda_i} m_{\lambda_i}} \right]_{LM},
\end{align}

where $\phi_{K_i}^{\ell_{\eta_i}, \ell_{\lambda_i}}(\theta_i)$ is a normalised hyperangular polynomial given by

\begin{align} \label{eq:hyper_polynomial}
\phi_{K_i}^{\ell_{\eta_i}, \ell_{\lambda_i}}(\theta_i) = N_{K_i}^{\ell_{\eta_i}, \ell_{\lambda_i}} \, (\sin \theta_i)^{\ell_{\eta_i}} (\cos \theta_i)^{\ell_{\lambda_i}} P_{n_i}^{(\ell_{\eta_i} + \frac{1}{2}, \ell_{\lambda_i} + \frac{1}{2})}(\cos 2\theta_i),
\end{align}

$P^{(p_i,q_i)}_{n_i}(x)$ is the Jacobi polynomial, with $p_i=\ell_{\eta_i} + \frac{1}{2}$ and $q_i=\ell_{\lambda_i} + \frac{1}{2}$. The normalisation constant $N_{K}^{\ell_{\eta_i}, \ell_{\lambda_i}} $ is given by

\begin{align}
N_{K_i}^{\ell_{\eta_i}, \ell_{\lambda_i}} = \sqrt{\frac{2(2n_i+\ell_{\eta_i}+\ell_{\lambda_i}+2) \, n_i! \, \Gamma(n_i+\ell_{\eta_i}+\ell_{\lambda_i}+2)}{\Gamma(n_i+\ell_{\eta_i}+\frac{3}{2}) \, \Gamma(n_i+\ell_{\lambda_i}+\frac{3}{2})}}
\end{align}
The hyperangular momentum $K_i$ is a nonnegative integer i.e. $K_i=0,1,2, \ldots$. The quantum numbers $K_i$, $ \ell_{\eta_i}$ and $\ell_{\lambda_i}$  are restricted by the relation $K_i - (\ell_{\eta_i} +\ell_{\lambda_i})=2n_i$, where $n_i=0,1,2,3, ...$.

The tensor product $\left[ Y^{\ell_{\eta_i} m_{\eta_i}} \otimes Y^{\ell_{\lambda_i} m_{\lambda_i}} \right]_{LM}$ represents the angular momentum coupling of two spherical harmonics to form a combined state with total angular momentum $L$ and magnetic quantum number $M$. In other words, one is given the orbital angular momentum $\ell_{\eta_i}$ with its projection $m_{\eta_i}$ and the orbital angular momentum $\ell_{\lambda_i}$ with its projection $m_{\lambda_i}$ to construct the state with total angular momentum $L$ and magnetic quantum number $M$. Explicitly, this coupled state can be written as summations of products of spherical harmonics over the magnetic quantum numbers:

\begin{align}\label{eq:clebsch}
\left[ Y^{\ell_{\eta_i} m_{\eta_i}} \otimes Y^{\ell_{\lambda_i} m_{\lambda_i}} \right]_{LM} = \sum_{m_{\eta_i},m_{\lambda_i}} C^{LM}_{\ell_{\eta_i}m_{\eta_i},\ell_{\lambda_i}m_{\lambda_i}} \, Y^{\ell_{\eta_i} m_{\eta_i}}(\nu_{\eta_i}, \omega_{\eta_i}) \, Y^{\ell_{\lambda_i} m_{\lambda_i}}(\nu_{\lambda_i}, \omega_{\lambda_i})
\end{align}

where $C^{LM}_{\ell_{\eta_i}m_{\eta_i},\ell_{\lambda_i}m_{\lambda_i}}=\langle \ell_{\eta_i} m_{\eta_i}, \ell_{\lambda_i} m_{\lambda_i} | LM \rangle$ are the Clebsch-Gordan coefficients.

\subsection{Further factorisation of the integral: Angular integration}

Of the five angles, \(V_i\) depends only on \(\theta_i\). Hence, the integration over the four spherical angles can be performed separately. The integral \(V^{\ell_{\eta_i},\ell_{\lambda_i};\,\ell_{\eta_i}',\ell_{\lambda_i}'}_{K_iK_n}(\rho)\) therefore becomes

\begin{equation}
V^{\ell_{\eta_i},\ell_{\lambda_i};\,\ell_{\eta_i}',\ell_{\lambda_i}'}_{K_iK_n}(\rho) = \int_0^{\pi/2} \Phi_{K_i}^{\ell_{\eta_i}, \ell_{\lambda_i}} (\theta_i) \, V_i(\rho \sin\theta_i) \, \Phi_{K_n}^{\ell_{\eta_i'}, \ell_{\lambda_i'}} (\theta_i) \, \sin^2\theta_i \, \cos^2\theta_i \, d\theta_i \times \mathcal{A},
\end{equation}

where the angular factor \(\mathcal{A}\) is:

\begin{align}
\mathcal{A}
&=
\sum_{m_{\eta_i},\,m_{\lambda_i}}
\sum_{m_{\eta_i}',\,m_{\lambda_i}'}
\left(
C_{\ell_{\eta_i} m_{\eta_i},\, \ell_{\lambda_i} m_{\lambda_i}}^{LM}
\right)^{*}
C_{\ell_{\eta_i}' m_{\eta_i}',\, \ell_{\lambda_i}' m_{\lambda_i}'}^{LM}
\nonumber \\
&\quad\times
\int
Y_{\ell_{\eta_i}}^{m_{\eta_i} *}(\Omega_{\eta_i})\,
Y_{\ell_{\eta_i}'}^{m_{\eta_i}'}(\Omega_{\eta_i})\,
d\Omega_{\eta_i}
\int
Y_{\ell_{\lambda_i}}^{m_{\lambda_i} *}(\Omega_{\lambda_i})\,
Y_{\ell_{\lambda_i}'}^{m_{\lambda_i}'}(\Omega_{\lambda_i})\,
d\Omega_{\lambda_i}.
\end{align}

The orthonormality of spherical harmonics is applied,

\begin{equation}
\int Y_{\ell}^{m*} Y_{\ell'}^{m'} d\Omega = \delta_{\ell \ell'} \delta_{m m'}.
\end{equation}

This gives rise to the following form of the angular factor

\begin{equation}
\mathcal{A}
=
\delta_{\ell_{\eta_i}\ell_{\eta_i}'}
\,
\delta_{\ell_{\lambda_i}\ell_{\lambda_i}'}
\sum_{m_{\eta_i},\,m_{\lambda_i}}
\left|
C_{\ell_{\eta_i} m_{\eta_i},\, \ell_{\lambda_i} m_{\lambda_i}}^{LM}
\right|^{2}
=
\delta_{\ell_{\eta_i}\ell_{\eta_i}'}
\,
\delta_{\ell_{\lambda_i}\ell_{\lambda_i}'} .
\end{equation}

Therefore, the integral vanishes unless $\ell_{\eta_i} = \ell_{\eta_i}'$ and $\ell_{\lambda_i} = \ell_{\lambda_i}'$ i.e. it is diagonal with respect to the \((\ell_{\eta}, \ell_{\lambda})\) quantum numbers. The diagonality shows that central potentials do not mix different orbital angular momentum components of the spectator or the interacting pair during the transformation between Jacobi sets. This is a consequence of the fact that central potentials have no angular dependence. Any mixing between different orbital configurations must arise from non-central interactions (e.g. tensor, spin–orbit, or spin–spin forces). The coupling integral may therefore be written as

\begin{equation}
V^{\ell_{\eta_i},\,\ell_{\lambda_i};\,\ell_{\eta_i}',\,\ell_{\lambda_i}'}_{K_iK_n}(\rho)
=
\delta_{\ell_{\eta_i}\ell_{\eta_i}'}
\,
\delta_{\ell_{\lambda_i}\ell_{\lambda_i}'}
\;
J^{\ell_{\eta_i},\,\ell_{\lambda_i}}_{K_iK_n}(\rho) .
\end{equation}

The reduced radial-hyperangular integral \(J^{\ell_{\eta_i},\,\ell_{\lambda_i}}_{K_iK_n}(\rho) \) is

\begin{equation}
\label{eq:reduced}
J^{\ell_{\eta_i},\,\ell_{\lambda_i}}_{K_iK_n}(\rho)
=
\int_{0}^{\pi/2}
\Phi^{\ell_{\eta_i},\,\ell_{\lambda_i}}_{K_i}(\theta_i)\,
V\!\left(\rho \sin\theta_i\right)\,
\Phi^{\ell_{\eta_i},\,\ell_{\lambda_i}}_{K_n}(\theta_i)\,
\sin^{2}\theta_i \cos^{2}\theta_i \,
d\theta_i .
\end{equation}

Substituting back, we obtain

\begin{equation}
V^{in}_{K_iK_n}(\rho) = \sum_{\ell_{\eta_i},\,\ell_{\lambda_i}}
RR\left\langle
\ell_{\eta_i} \ell_{\lambda_i}
\,\middle|\,
\ell_{\eta_n} \ell_{\lambda_n}
\right\rangle_{\mathrm{K_nL}} \cdot J_{K_iK_n}^{\ell_{\eta_i}, \ell_{\lambda_i}} (\rho).
\end{equation}

The coupling potential is now factorized into the Raynal-Revai coefficient and the integral \(J_{K_iK_n}^{\ell_{\eta_i}, \ell_{\lambda_i}} (\rho)\) over the hyperangle \(\theta_i\), which contains the potential \(V_i(\rho \sin\theta_i)\). This separation demonstrates that the complexity of channel coupling in three--body systems arises not from the coordinate transformation between Jacobi sets, but from the hyperangular dependence of the interaction potential. The Raynal--Revai coefficient is a geometric factor since it arises purely from the coordinate transformation between different Jacobi sets. The integral \(J_{K_iK_n}^{\ell_{\eta_i}, \ell_{\lambda_i}} (\rho)\) over the hyperangle \(\theta_i\) is a dynamical factor since it carries the interaction potential. Therefore, the coupling potential is factorized into a geometric part and a dynamical part. The asymptotic behaviour of the Raynal--Revai coefficients at large hyperangular momenta has been explicitly developed in \citep{pup2009}. Numerous algorithms have been proposed for computing the Raynal--Revai coefficients \citep{you1987, kha1999, ers2016}. The asymptotic properties of coupling matrix elements for central potentials have been rigorously analyzed within the adiabatic hyperspherical framework by \cite{kvi1991}, who demonstrated that for a wide class of short-range interactions, the coupling strength exhibits decay characteristics determined primarily by the angular momentum of the interacting pair. It was demonstrated numerically in \cite{kle1990} that coupling matrix elements for short-range potentials decay algebraically in the hyperradius in the large-$\rho$ limit,

\section{Results and discussion}

In this section, we analyse the behaviour of the reduced hyperangular coupling integral $J^{\ell_{\eta_i},\ell_{\lambda_i}}_{K_iK_n}(\rho)$ for different classes of two--body potentials. Particular emphasis is placed on the asymptotic behaviour as the hyperradius $\rho\to\infty$, since this governs the strength of channel coupling and the convergence properties of hyperspherical expansions in three--body calculations. Three short--range potentials (Gaussian, Yukawa and Woods--Saxon potentials) and one long--range interaction
(Coulomb potential) are considered. For ease of notation, $J^{\ell_{\eta_i},\ell_{\lambda_i}}_{K_iK_n}(\rho)$ shall be written as $J^{\ell_{\eta_i},\ell_{\lambda_i}}_{KK'}(\rho)$, where $K=\ell_{\eta_i} + \ell_{\lambda_i} + n$ and $K'=\ell_{\eta_i} + \ell_{\lambda_i} + n'$.   

\subsection{Short--range potentials}

\subsubsection{Gaussian potential}

A Gaussian two--body interaction considered is

\begin{equation}
V(r) = V_0\, e^{-r^2/a_0^2},
\end{equation}

where $r$ is  the interparticle distance, $a_0$ is the range parameter and $V_0$ is the strength (or depth). If $V_0 < 0$ the potential is attractive and if $V_0>0$ the potential is repulsive. In hyperspherical coordinates, this potential takes the form

\begin{equation}
V_i(\rho\sin\theta_i) = V_0 \exp\!\left[-\frac{\rho^2}{a_0^2}\sin^2\theta_i\right].
\end{equation}

Substituting this into Eq.~\eqref{eq:reduced} yields

\begin{equation}
J^{\ell_{\eta_i},\ell_{\lambda_i}}_{KK'}(\rho)
=
V_0
\int_0^{\pi/2}
\Phi^{\ell_{\eta_i},\,\ell_{\lambda_i}}_{K}(\theta_i) \Phi^{\ell_{\eta_i},\,\ell_{\lambda_i}}_{K'}(\theta_i)\,
e^{-(\rho^2/a_0^2)\sin^2\theta_i}\,
\sin^2\theta_i\cos^2\theta_i\, d\theta_i .
\end{equation}

Writing the hyperangular functions explicitly, one obtains

\begin{align}
J^{\ell_{\eta_i},\ell_{\lambda_i}}_{KK'}(\rho)&=
V_0 N_{K}^{\ell_{\eta_i}, \ell_{\lambda_i}} N_{K'}^{\ell_{\eta_i}, \ell_{\lambda_i}}
\int_0^{\pi/2}
(\sin\theta)^{2\ell_{\eta_i}+2}(\cos\theta)^{2\ell_{\lambda_i}+2}
\nonumber\\
&\quad\times
P_{n}^{(\ell_{\eta_i} + \frac{1}{2}, \ell_{\lambda_i} + \frac{1}{2})}(\cos 2\theta_i)
P_{n'}^{(\ell_{\eta_i} + \frac{1}{2}, \ell_{\lambda_i} + \frac{1}{2})}(\cos 2\theta_i)
e^{-(\rho^2/a_0^2)\sin^2\theta_i}
\, d\theta .
\end{align}

Consider the asymptotic region $\rho \to \infty$. For a short-range two-body interaction, the potential $V_i(\eta_i)$ is non-negligible only when the interacting pair separation $\eta_i$ remains within the finite interaction range. Since \(\eta_i = \rho \sin\theta_i\), in the asymptotic region $\rho \to \infty$, one must have $\theta_i \to 0$ for $\eta_i$ to remain within the finite interaction range. In this region,

\begin{align}
   \sin\theta_i \simeq \theta_i, \qquad \cos\theta_i \simeq 1, 
\end{align}

For $\theta_i \to 0$ the Jacobi polynomials approach finite constants $P_{n_i}^{(\ell_{\eta_i} + \frac{1}{2}, \ell_{\lambda_i} + \frac{1}{2})}(1)$ given by 

\begin{align}
P^{(p_i,q_i)}_{n}(1) = \binom{n+p_i}{n}
\end{align}

where $p_i=\ell_{\eta_i} + \frac{1}{2}$ and $q_i=\ell_{\lambda_i} + \frac{1}{2}$. Therefore, in the asymptotic region $\rho \to \infty$, the coupling integral becomes

\begin{equation}
J^{\ell_{\eta_i},\ell_{\lambda_i}}_{KK'}(\rho)
\sim
V_0 N_{K}^{\ell_{\eta_i}, \ell_{\lambda_i}} N_{K'}^{\ell_{\eta_i}, \ell_{\lambda_i}} P^{(p_i,q_i)}_{n}(1) P^{(p_i,q_i)}_{n'}(1)
\int_0^\infty
{\theta_i}^{2\ell_{\eta_i}+2}
e^{-(\rho^2/a_0^2){\theta_i}^2}
\, d\theta_i.
\end{equation}

Evaluating the Gaussian moment integral yields 

\begin{align}
J^{\ell_{\eta_i},\ell_{\lambda_i}}_{KK'}(\rho)
\sim
V_0 N_{K}^{\ell_{\eta_i}, \ell_{\lambda_i}} N_{K'}^{\ell_{\eta_i}, \ell_{\lambda_i}} P^{(p_i,q_i)}_{n}(1) P^{(p_i,q_i)}_{n'}(1)\frac{1}{2}\,
\Gamma\!\left(\ell_{\eta_i}+\frac{3}{2}\right)
\left(\frac{a_0}{\rho}\right)^{2\ell_{\eta_i}+3}
\end{align}

The asymptotic scaling is therefore

\begin{equation}
J^{\ell_{\eta_i},\ell_{\lambda_i}}_{KK'}(\rho)
\;\propto\;
\rho^{-(2\ell_{\eta_i}+3)},
\qquad \rho\to\infty .
\end{equation}

This shows that Gaussian potentials produce extremely rapid algebraic suppression of channel couplings i.e. different hyperspherical channels become effectively uncoupled at large distances. Only a finite number of channels contribute significantly in the asymptotic region.  This asymptotic decoupling of hyperspherical channels for short-range interactions has been systematically analysed in \cite{nie2001}, in which the large-distance behaviour of the non-diagonal coupling terms is derived within the adiabatic hyperspherical method.

\subsubsection{Yukawa Potential}

Another common potential in nuclear physics is the Yukawa potential, which is also sometimes called a screened Coulomb potential. It is given by

\begin{equation}
V(r) = V_0 \frac{e^{-\mu r}}{r},
\end{equation}

where $r$ is the distance between the interacting particles, $V_0$ is the strength (coupling constant) of the interaction and $\mu$ is the inverse range parameter (or screening parameter) of the interaction. If $V_0 < 0$ the potential is attractive and if $V_0>0$ the potential is repulsive. This potential behaves like a Coulomb potential at short distances when $\mu r \ll 1$. Many central potentials are combinations of Yukawa functions. For example, the Malfliet–Tjon potential for the nucleon-nucleon interaction is a sum of two Yukawa terms, one representing short-range repulsion and the other intermediate-range attraction.

In hyperspherical coordinates this potential is given by

\begin{equation}
V_i(\rho\sin\theta_i)
=
\frac{V_0}{\rho}
\frac{e^{-\mu\rho\sin\theta_i}}{\sin\theta_i}.
\end{equation}

Substituting into Eq.~\eqref{eq:reduced} gives

\begin{equation}
J^{\ell_{\eta_i},\ell_{\lambda_i}}_{KK'}(\rho)
=
\frac{V_0}{\rho}
\int_0^{\pi/2}
\Phi^{\ell_{\eta_i},\,\ell_{\lambda_i}}_{K}(\theta_i) \Phi^{\ell_{\eta_i},\,\ell_{\lambda_i}}_{K'}(\theta_i)\,
\sin\theta_i\cos^2\theta_i\,
e^{-\mu\rho\sin\theta_i}\,
d\theta_i .
\end{equation}

Explicitly,

\begin{align}
J^{\ell_{\eta_i},\ell_{\lambda_i}}_{KK'}(\rho)
&=
\frac{V_0}{\rho} N_{K}^{\ell_{\eta_i}, \ell_{\lambda_i}} N_{K'}^{\ell_{\eta_i}, \ell_{\lambda_i}}
\int_0^{\pi/2}
(\sin\theta_i)^{2\ell_{\eta_i}+1}(\cos\theta_i)^{2\ell_{\lambda_i}+2}
\nonumber\\
&\quad\times
P_{n}^{(\ell_{\eta_i} + \frac{1}{2}, \ell_{\lambda_i} + \frac{1}{2})}(\cos 2\theta_i)
P_{n'}^{(\ell_{\eta_i} + \frac{1}{2}, \ell_{\lambda_i} + \frac{1}{2})}(\cos 2\theta_i)
e^{-\mu\rho\sin\theta}\, d\theta_i .
\end{align}

As explained earlier, for $\rho\to\infty$, the dominant contribution arises from small hyperangles i.e. $\theta_i \to 0$. Therefore $\sin\theta_i\simeq\theta_i$ and $\cos\theta_i \to 1$, with the Jacobi polynomials taking on the same finite values. The integral reduces to

\begin{equation}
J^{\ell_{\eta_i},\ell_{\lambda_i}}_{KK'}(\rho)
\sim
V_0N_{K}^{\ell_{\eta_i}, \ell_{\lambda_i}} N_{K'}^{\ell_{\eta_i}, \ell_{\lambda_i}}
P_{n}^{p_i,q_i}(1)
P_{n'}^{p_i,q_i}(1)
\frac{1}{\rho}
\int_0^\infty
{\theta_i}^{2\ell_{\eta_i}+1}
e^{-\mu\rho\theta_i}\,
d\theta_i .
\end{equation}

Evaluating this integral yields

\begin{align}
 J^{\ell_{\eta_i},\ell_{\lambda_i}}_{KK'}(\rho)
\sim
V_0N_{K}^{\ell_{\eta_i}, \ell_{\lambda_i}} N_{K'}^{\ell_{\eta_i}, \ell_{\lambda_i}}
P_{n}^{p_i,q_i}(1)
P_{n'}^{p_i,q_i}(1)
\frac{\Gamma(2\ell_{\eta_i}+2)}{\mu^{2\ell_{\eta_i}+2}\,\rho^{2\ell_{\eta_i}+3}} .   
\end{align}

Its asymptotic behaviour is therefore given by

\begin{equation}
J^{\ell_{\eta_i},\ell_{\lambda_i}}_{KK'}(\rho)
\;\propto\;
\rho^{-(2\ell_{\eta_i}+3)},
\qquad \rho\to\infty .
\end{equation}

This shows that Yukawa potentials produce the same power law asymptotic behaviour as Gaussians. This reveals extremely rapid algebraic suppression of channel couplings, just like Gaussian potentials.

\subsubsection{Woods--Saxon potential}

The Woods--Saxon (WS) potential is a widely used phenomenological interaction in nuclear physics that is used to model the average (mean-field) interaction felt by a nucleon inside the nucleus. The Woods–Saxon potential effectively represents the mean field generated by all nucleons, rather than a fundamental two-body force. It is given by

\begin{align}
    V(r) = -\frac{V_{0}}{1 + \exp\!\left(\frac{r-R}{a}\right)},
\end{align}

where $r$ is the distance of the nucleon from the centre of the nucleus, \(V_{0}\) is the depth, \(R\) the radius parameter, and \(a\) the surface diffuseness.  If \(V_{0} >0\) the force is attractive and if \(V_{0} <0\) it is repulsive. 

In hyperspherical coordinates, the Woods--Saxon potential becomes

\begin{align}
    V_i(\rho\sin\theta_i) = -\frac{V_{0}}{1 + \exp\!\left(\frac{\rho\sin\theta_i - R}{a}\right)} .
\end{align}

Substituting this into the reduced hyperangular coupling integral, Eq.~\eqref{eq:reduced}, one arrives at

\begin{equation}
J^{\ell_{\eta_i},\ell_{\lambda_i}}_{KK'}(\rho)
= -V_{0}\int_{0}^{\pi/2} \Phi^{\ell_{\eta_i},\,\ell_{\lambda_i}}_{K}(\theta_i) \Phi^{\ell_{\eta_i},\,\ell_{\lambda_i}}_{K'}(\theta_i)\,
\frac{\sin^{2}\theta_i\cos^{2}\theta_i}{\,1+\exp\!\left(\frac{\rho\sin\theta_i-R}{a}\right)} \, d\theta_i .
\end{equation}

Writing the hyperangular functions explicitly yields

\begin{align}
J^{\ell_{\eta_i},\ell_{\lambda_i}}_{KK'}(\rho)
&= -V_{0} N_{K}^{\ell_{\eta_i}, \ell_{\lambda_i}} N_{K'}^{\ell_{\eta_i}, \ell_{\lambda_i}}
\int_{0}^{\pi/2} (\sin\theta_i)^{2\ell_{\eta_i}+2}(\cos\theta_i)^{2\ell_{\lambda_i}+2} \nonumber \\
&\quad \times 
P_{n}^{(\ell_{\eta_i} + \frac{1}{2}, \ell_{\lambda_i} + \frac{1}{2})}(\cos 2\theta_i)
P_{n'}^{(\ell_{\eta_i} + \frac{1}{2}, \ell_{\lambda_i} + \frac{1}{2})}(\cos 2\theta_i)
\frac{d\theta_i}{1+\exp\!\left(\frac{\rho\sin\theta_i-R}{a}\right)} .
\end{align}

As the hyperradius \(\rho\) grows to infinity, $\theta \to 0$. Therefore \(\rho\sin\theta_i \sim \rho \theta_i\). In this regime, the exponential in the denominator dominates whenever \(\rho \theta_i \gg R\).  One may approximate as follows:

\begin{align}
    1 + \exp\!\left(\frac{\rho\theta_i-R}{a}\right) \simeq 
\exp\!\left(\frac{\rho\theta_i-R}{a}\right) ,
\end{align}

Therefore,

\begin{align}
 \frac{1}{1+\exp\!\left(\frac{\rho\sin\theta_i-R}{a}\right)} \simeq 
\exp\!\left(-\frac{\rho\theta_i-R}{a}\right)
= e^{R/a}\,e^{-\rho\theta_i/a}.   
\end{align}

This shows that in the asymptotic region, the Woods–Saxon potential effectively behaves like an exponentially damped short-range interaction in the hyperangle. Inserting these approximations into the integral yields

\begin{equation}
J^{\ell_{\eta_i},\ell_{\lambda_i}}_{KK'}(\rho)
\simeq -V_{0}\,N_{K}^{\ell_{\eta_i}, \ell_{\lambda_i}} N_{K'}^{\ell_{\eta_i}, \ell_{\lambda_i}} P^{(p_i,q_i)}_{n}(1) P^{(p_i,q_i)}_{n'}(1)
\,e^{R/a}\int_{0}^{\infty} {\theta_i}^{2\ell_{\eta_i}+2}\,
e^{-\rho\theta_i/a}\, d\theta_i .
\end{equation}

Evaluating this integral, one arrives at 

\begin{equation}
J^{\ell_{\eta_i},\ell_{\lambda_i}}_{KK'}(\rho)
\sim -V_{0}\,N_{K_i}^{\ell_{\eta_i}, \ell_{\lambda_i}} N_{K'_i}^{\ell_{\eta_i}, \ell_{\lambda_i}} P^{(p_i,q_i)}_{n}(1) P^{(p_i,q_i)}_{n'}(1)
\,e^{R/a} \,
\left(\frac{a}{\rho}\right)^{\!2\ell_{\eta_i}+3}
\Gamma\!\bigl(2\ell_{\eta_i}+3\bigr)
\end{equation}

This reveals the following asymptotic behaviour for the Woods--Saxon potential:

\begin{align}
    J^{\ell_{\eta_i},\ell_{\lambda_i}}_{KK'}(\rho)\;\propto\;
\rho^{-(2\ell_{\eta_i}+3)},\qquad \rho\to\infty\
\end{align}

Like the Gaussian and Yukawa interactions, the Woods--Saxon potential produces algebraically suppressed channel couplings at large hyperradius. The decay exponent \(2\ell_{\eta_i}+3\) is identical to that of the Gaussian case, indicating that the asymptotic decoupling efficiency is governed primarily by the orbital angular momentum \(\ell_{\eta_i}\) of the interacting pair, not by the detailed shape of the short-range potential.  The prefactor, however, contains the diffuseness \(a\) and radius \(R\) explicitly through the factor \(e^{R/a}a^{2\ell_{\eta_i}+3}\). This reflects the fact that a larger diffuseness extends the effective range of the potential, slightly delaying the onset of the asymptotic power-law regime compared with a Gaussian of similar range. Nevertheless, the ultimate algebraic decay guarantees that hyperspherical channels eventually decouple, ensuring convergence of the hyperspherical harmonic expansion for three-body systems interacting through Woods--Saxon forces.

\subsection{Long--range potential: Coulomb potential}

The Coulomb potential between two point particles with charges $Z_je$ and $Z_ke$, separated by a distance $r$, the Coulomb potential is

\begin{equation}
V(r) = \frac{Z_j Z_k e^2}{r},
\end{equation}

where $e$ is the elementary charge. In hyperspherical coordinates, this potential becomes

\begin{equation}
V_i(\rho\sin\theta_i) = \frac{Z_j Z_k e^2}{\rho\sin\theta_i}.
\end{equation}

Substituting into Eq.~\eqref{eq:reduced} yields

\begin{equation}
J^{\ell_{\eta_i},\ell_{\lambda_i}}_{KK'}(\rho)
=
\frac{Z_j Z_k e^2}{\rho}
\int_0^{\pi/2}
\Phi^{\ell_{\eta_i},\,\ell_{\lambda_i}}_{K}(\theta_i) \Phi^{\ell_{\eta_i},\,\ell_{\lambda_i}}_{K'}(\theta_i)\,
\sin\theta_i\cos^2\theta_i\, d\theta_i .
\end{equation}

The integrand is independent of $\rho$, unlike the cases of Gaussian, Yukawa and Woods--Saxon potentials. Therefore, one may define a purely geometric Coulomb matrix

\begin{equation}
C^{\ell_{\eta_i},\ell_{\lambda_i}}_{KK'}
=
\int_0^{\pi/2}
\Phi^{\ell_{\eta_i},\,\ell_{\lambda_i}}_{K}(\theta_i) \Phi^{\ell_{\eta_i},\,\ell_{\lambda_i}}_{K'}(\theta_i)\,
\sin\theta_i\cos^2\theta_i\, d\theta_i ,
\end{equation}

The coupling potential therefore factorizes as

\begin{equation}
J^{\ell_{\eta_i},\ell_{\lambda_i}}_{KK'}
=
\frac{Z_j Z_k e^2}{\rho}\,
C^{\ell_{\eta_i},\ell_{\lambda_i}}_{KK'} .
\end{equation}

It may immediately be observed that the Coulomb potential produces channel couplings that decay only as $1/\rho$, and this power law has no dependence on $\ell_{\eta_i}$, as is the case with the short-range potentials studied. As a result, hyperspherical channels remain coupled at arbitrarily large hyperradii. This explains the slow convergence of hyperspherical harmonic expansions in charged three--body systems. The slow $1/\rho$ decay of Coulomb couplings, derived analytically here, is consistent with the observations in \cite{zho1993, gus1990, laz2006}.

\section{Conclusion}

In this paper, we have systematically analysed the asymptotic behaviour of channel-coupling potentials in three-body systems using the hyperspherical harmonics expansion method. By expressing the coupling in terms of Raynal-Revai coefficients and a reduced hyperangular integral, we demonstrated that the coupling potential factorizes into a geometric part (from coordinate transformations) and a dynamical part (from the two-body interaction). For Gaussian, Yukawa, and Woods-Saxon potentials, which are short-range potentials, we derived explicit asymptotic scaling laws. These laws show that the coupling strength decays algebraically as $\rho^{-(2\ell_{\eta_i}+3)}$ for large hyperradius $\rho$. This behaviour, governed by the orbital angular momentum $\ell_{\eta_i}$ of the interacting pair, ensures rapid asymptotic decoupling of hyperspherical channels and supports the practical convergence of hyperspherical expansions.

In contrast, the long-range Coulomb potential produces couplings that decay only as $1/\rho$, leading to persistent interchannel coupling at arbitrarily large distances. This result explains the well-known slow convergence of hyperspherical harmonic expansions in charged three-body systems and underscores the fundamental difference between short- and long-range interactions in many-body dynamics.

Our findings provide a clear analytic foundation for understanding channel coupling in three-body problems and offer practical guidance for truncating hyperspherical expansions in nuclear physics calculations. The explicit asymptotic forms derived here can be used to assess the importance of channel couplings at large distances and to develop more efficient numerical methods for solving three-body Faddeev equations in both bound and scattering states.

\bibliographystyle{unsrtnat}
\bibliography{coupling_potentials_meoto}
\end{document}